# Soft X-Ray Single-Photon Detection With Superconducting Tantalum Nitride and Niobium Nanowires

Kevin Inderbitzin, Andreas Engel, and Andreas Schilling

We have fabricated ultrafast dark count-free soft X-ray single-photon detectors (X-SNSPDs) from TaN with various conduction path widths, and we compare their properties with corresponding data from a Nb X-SNSPD. The TaN X-SNSPDs offer an improved detector performance regarding device detection efficiency, latching and pulse amplitudes. Wide conduction paths allow for a certain energy-resolving capability in contrast to narrow TaN conduction paths. However, wide paths also limit the detection efficiency at low temperatures, which can be explained within a hot-spot model.

## I. Introduction

X-RAY superconducting nanowire single-photon detectors (X-SNSPDs) are good candidates [1] for applications where very high count rates, precise timing, negligible dark counts and response in the soft X-ray energy range are required. Potential applications cover experiments with free-electron lasers, synchrotron X-ray sources and hot plasmas as in nuclear fusion experiments.

While superconducting nanowire single-photon detectors (SNSPDs) have been shown to be able to detect infrared and optical photons [2] as well as molecules with keV-energies [3] for time-of-flight mass spectrometry, we have very recently also demonstrated the successful operation of an X-SNSPD [1]. As Perez de Lara et al. [4] had shown, X-ray photon detection in conventional thin-film SNSPDs (with a typical thickness $\approx 5$ nm) mainly occurs by photon absorption in the substrate and subsequent energy diffusion to the superconducting meander structure. Since this could hamper the timing jitter of a single-photon detector, we had fabricated and studied an SNSPD from a 100 nm thick niobium film (a so called X-SNSPD), in order to promote the direct X-ray photon absorption in the superconducting structure. We achieved dark count-free single-photon detection of keV-photons in continuous mode for reduced bias currents down to 0.4% of $I_c$ with this detector, but with latching appearing above 5.5% of $I_c$. Very interestingly, the signal amplitude distribution depended significantly on the photon energy spectrum, which may allow for a certain energy resolution.

Dated December 12, 2012

This research received support from the Swiss National Science Foundation Grant No. 200021_135504/1.

K. Inderbitzin, A. Engel, and A. Schilling are with the Physics Institute, University of Zurich, 8057 Zurich, Switzerland (e-mail: kevin.inderbitzin@physik.uzh.ch).

We have now also developed X-SNSPDs from a 100 nm thick tantalum nitride film in order to improve several detector properties as compared to the Nb X-SNSPD [1]. Firstly, TaN offers a slightly better soft X-ray absorptance than niobium. Secondly, TaN has an about two orders of magnitude larger resistivity than Nb, which can lead to larger resistances of the normal-conducting domains that emerge after photon detection. This should result in larger signal amplitudes and a better signal-to-noise ratio. In addition, larger domain resistances are expected to reduce problems with latching [5], thus allowing for continuous photon-counting at larger reduced bias currents. Thirdly, a 100 nm thick TaN film has an effective superconducting penetration depth large enough to guarantee for a homogeneous current distribution even in several $\mu m$-wide conduction paths, which is essential for a good device detection efficiency (DDE). However, the energy-dependence of the signal amplitude distribution as we observed it in the Nb X-SNSPD [1] is expected to be reduced by the larger domain resistances. In order to partially regain this energy-dependence with smaller domain resistances, we also fabricated and examined a TaN X-SNSPD with very wide conduction paths.

In this paper we report on experiments with TaN X-SNSPDs with an improved detector performance as compared to Nb X-SNSPDs regarding DDE, latching and pulse amplitudes. The energy-dependence of the signal amplitude in the keV-range is weak for $\approx 250$ nm wide TaN conduction paths, but it is pronounced for $\approx 1.6 \mu m$ wide paths.

## II. Device Fabrication

A tantalum nitride film of thickness $d \approx 100$ nm was grown at $T = 700\,°C$ by DC reactive magnetron sputtering in an Ar/N$_2$ atmosphere on an R-plane cut sapphire substrate. The as-grown film showed a critical temperature $T_c = 8.0$ K and a resistivity of $\rho_{TaN}(10.3\text{ K}) \approx 500\,\mu\Omega cm$, which is two orders of magnitude larger than for a 100 nm thick Nb film [1] with $\rho_{Nb}(10\text{ K}) \approx 4\,\mu\Omega cm$ right above $T_c$. Two different X-SNSPDs were fabricated using electron-beam lithography using ZEP520A resist and reactive ion etching [6]. Fig. 1 shows an optical image of both detectors with the most relevant properties summarized in Table I. Detector TaN-A has a uniform conduction path width $w$ of $\approx 250$ nm and a

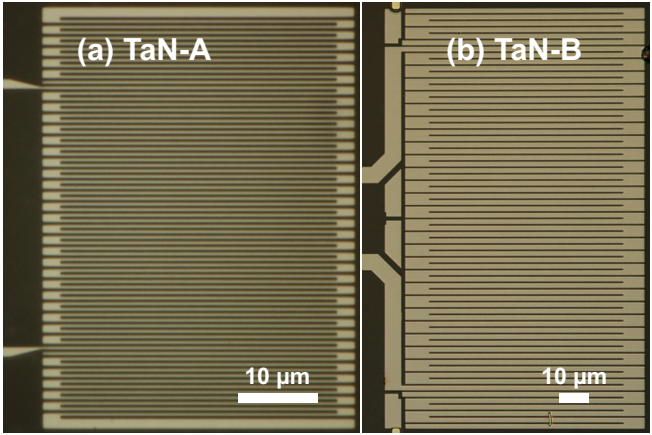

Fig. 1. Optical images of the examined X-SNSPDs from 100 nm thick TaN: (a) Detector TaN-A with a conduction path width of $\approx 250\,\text{nm}$ and a filling factor $\approx 35\%$. (b) Detector TaN-B has a much wider conduction path width $\approx 1.6\,\mu\text{m}$ and a filling factor $\approx 75\%$. The large effective penetration depth $\lambda_{\text{eff}}$ ensures a homogeneous current distribution in these conduction paths.

TABLE I
RELEVANT DETECTOR PROPERTIES

| Detector property | Detector TaN-A | Detector TaN-B |
|---|---|---|
| Conduction path width $w$ | $\approx 250\,\text{nm}$ | $\approx 1.6\,\mu\text{m}$ |
| Conduction path length $l$ | 1.67 mm | 3.59 mm |
| Active detector area | 35 x 33 $\mu\text{m}^2$ | 75 x 119 $\mu\text{m}^2$ |
| Filling factor | $\approx 35\%$ | $\approx 75\%$ |
| Critical temperature $T_c$ | 6.7 K | 7.0 K |
| Critical current $I_c$ | 105 µA (1.85 K) | 860 µA (1.85 K) |
|  |  | 430 µA (4 K) |
|  |  | 155 µA (5 K) |
| Calculated kinetic inductance $L_{\text{kin}}$ | 180 nH | 50 nH |
| Latching bias current (1.85 K, $V_A = 49.9$ kV) | > 52% $I_c$ | > 32% $I_c$ |
| Pulse rise time $T_R$ | 750 ps ± 60 ps | 910 ps ± 120 ps |

Relevant detector properties of the two X-SNSPDs fabricated from a 100 nm thick TaN film. Detector TaN-B has a conduction path width $w$ almost five times wider than detector TaN-A.

filling factor of 35%. After fabrication, it showed a critical temperature $T_c = 6.7\,\text{K}$. Detector TaN-B has a much wider conduction-path width of $\approx 1.6\,\mu\text{m}$, resulting in a higher filling factor of 75%. Its critical temperature is slightly higher than for detector TaN-A, with $T_c = 7.0\,\text{K}$.

These detector geometries ensure that the kinetic inductances are small enough for ultrafast recovery times [1], [7], and are at the same time large enough to reduce problems with latching as observed in [1], [8]. In the dirty limit, the bulk penetration depth can be calculated [9] from $T_c$ and the normal-state resistivity $\rho_N$ to $\lambda_{\text{bulk}} \approx 0.9 - 1.0\,\mu\text{m}$ for $T = 1.85\,\text{K}$, and $\lambda_{\text{bulk}} \approx 1.2 - 1.4\,\mu\text{m}$ for $T = 5\,\text{K}$ if its temperature dependence is considered [9]. As $d << \lambda_{\text{bulk}}$, the effective penetration depth can be determined [10] from $\lambda_{\text{eff}} = 2\lambda_{\text{bulk}}^2 / d \approx 17 - 22\,\mu\text{m}$ at $T = 1.85\,\text{K}$, which ensures a homogeneous current distribution in both detectors. The kinetic inductances can be determined [11] to $L_{\text{kin}} = \mu_0 \lambda_{\text{eff}} l / w \approx 180\,\text{nH}$ and 50 nH for detector TaN-A and TaN-B, respectively.

### III. EXPERIMENTAL SETUP

The experimental setup is identical to the one used for the characterization of the Nb X-SNSPD in [1]. Only one detector was characterized at a time. The detector signal was transmitted through a $50\,\Omega$ impedance signal line to a cryogenic amplifier at the 4 K-stage, and then further to a second amplifier at room temperature before fed into a 3.5 GHz digital oscilloscope. The amplifier chain had an effective bandwidth of about 40 MHz to 1.9 GHz. A quasi-constant-current bias [12] was applied and was passed through a series of low-pass filters.

The detectors were irradiated with X-ray photons through a $100\,\mu\text{m}$ thin window of polyimide (Kapton) at the cryostat. A second window of $1\,\mu\text{m}$ thin aluminum was installed at the 4 K-heat shield to minimize thermal radiation. Two qualitatively different X-ray sources were used: a tungsten-target X-ray tube with a maximum acceleration voltage $V_A = 49.9\,\text{kV}$, and a radioactive Fe-55 isotope source with an activity of 3.7 GBq, which emits photons with well defined energies around 6 keV. The homogeneous irradiation of the X-SNSPD by both sources was ensured.

### IV. EXPERIMENTAL RESULTS AND DISCUSSION

#### A. Voltage Signal

Upon X-ray irradiation we recorded voltage signal pulses with a shape qualitatively very similar to the pulse shapes recorded for the Nb X-SNSPD in [1] and the TaN thin-film SNSPD in [13]. The rise time $T_R$, defined as the time span between 10% and 90% of the signal maximum, was determined to be 750 ps ± 60 ps and 910 ps ± 120 ps for detector TaN-A and TaN-B, respectively. This is significantly longer than reported for the Nb X-SNSPD (i.e. 250 ps ± 70 ps), which we explain with the larger kinetic inductances of the TaN detectors. The pulse fall time is on the order of 10 ns for both detectors.

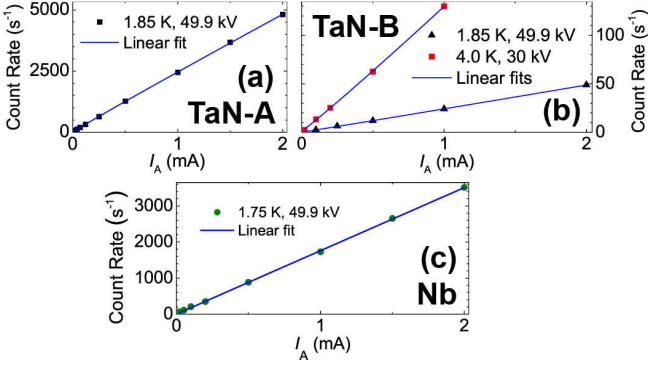

Fig. 2. Photon count rates as functions of $I_A$ of the detectors (a) TaN-A and (b) TaN-B and (c) the Nb X-SNSPD from [1], showing a linear dependence and therefore demonstrating single-photon detection.

### B. Single-Photon Detection

Fig. 2 shows the count-rate dependences on the X-ray photon flux (which was varied by the X-ray tube anode current $I_A$) at $V_A = 49.9\,\text{kV}$ for the detectors TaN-A, TaN-B and the Nb X-SNSPD [1]. For detector TaN-B, measurements at $T = 4.0\,\text{K}$ are also shown. These dependences show a linear behavior, and we therefore conclude that these X-SNSPDs detect in single-photon mode [2].

### C. Latching

The critical currents at $1.85\,\text{K}$ were determined to $I_c = 105\,\mu\text{A}$ and $860\,\mu\text{A}$ for detector TaN-A and TaN-B, respectively (Table I also gives the values for $4.0\,\text{K}$ and $5.0\,\text{K}$ for detector TaN-B). Fig. 3 shows the bias current-dependence of the count rate at $V_A = 49.9\,\text{kV}$ at $T = 1.85\,\text{K}$ for detector TaN-A and TaN-B. They latch for reduced bias currents $I_b/I_c > 52\%$ and $> 32\%$, respectively, which means that the devices cannot be continuously used for X-ray photon detection above these currents [5]. In these X-SNSPDs, latching occurs for much higher reduced bias currents than previously reported for the Nb X-SNSPD [1], where latching was observed for $I_b/I_c > 5.5\%$. This improvement can be attributed to the larger domain resistances in the TaN X-SNSPDs and possibly the higher kinetic inductances. For the same reasons, detector TaN-B latches for lower reduced bias currents than detector TaN-A, as its wider conduction paths lead to lower domain resistances, and its kinetic inductance is also lower.

### D. Hot-Spot Model

A hot-spot model used by [1], [14], [15] can be used to estimate the hot-spot radius after X-ray photon absorption. It is assumed that the primary excitations of low keV-photons are created in a small volume centered at the absorption point of the photon, thereby creating a hot-spot resembling a sphere in which superconductivity is destroyed. The bias current is therefore forced to flow in a smaller cross-section where it can exceed $I_c$ for large enough bias currents, which leads to the formation of a normal-conducting domain. By using the density-of-states at the Fermi-energy in TaN from [13] to determine the electronic specific heat, the hot-spot radius at $T = 1.85\,\text{K}$ for a $6\,\text{keV}$ and a $50\,\text{keV}$ photon can be estimated to 180-190 nm and 370-380 nm (according to [1]), respectively. In this model, the hot-spot radius increases only slightly when increasing the temperature to $T = 5\,\text{K}$, namely to 210 nm and 420 nm for $6\,\text{keV}$ and $50\,\text{keV}$ photons, respectively, assuming $T_c \approx 7\,\text{K}$.

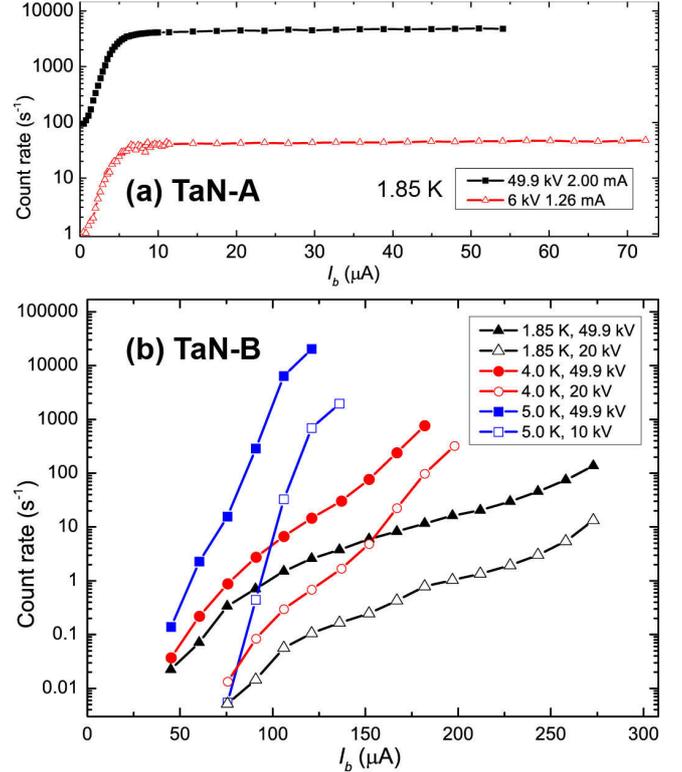

Fig. 3. Bias-current dependence of the photon count rate for different acceleration voltages, for the detectors (a) TaN-A and (b) TaN-B. For detector TaN-B, we show measurements at different temperatures.

### E. Bias-Current Dependence

Fig. 3(a) shows a plateau in the bias-dependent count rate for detector TaN-A down to $I_b < 7\,\mu\text{A} \approx 7\% \, I_c$, which can be explained with the fact that the calculated hot-spot diameter for a $6\,\text{keV}$ photon is larger than the conduction path width, and therefore a hot-spot is expected to cover the whole cross-section, independent of the bias current. Hence the decrease of the count rate for $I_b < 7\% \, I_c$ is not necessarily due to a small hot-spot radius for low-keV photons, but it might be attributed to a decreasing signal amplitude with a smaller bias current, while the trigger level is kept constant.

For detector TaN-B there is no plateau in the count rate dependence on $I_b$ (shown in Fig. 3(b) for different temperatures and acceleration voltages). This is in agreement with the expectations from the hot-spot model: as the hot-spot diameter is smaller than the conduction path width even for $50\,\text{keV}$ photons, a bias current above a non-zero threshold [14] must be applied in order to detect photons efficiently. It has been demonstrated that this bias threshold can depend on the energy of detected β-particles [16] and keV ions [15] and also on the operation temperature [16]. At

$T = 1.85\,\text{K}$, however, we expect that the energy-dependent bias thresholds are larger than $460\,\mu\text{A}$ (assuming a cylindrical hot-spot [14]), which is already in the latching regime. This explains why the count rates at $T = 1.85\,\text{K}$ remain low even for high X-ray intensities (detector TaN-A shows $> 4 \cdot 10^3$ cps at the same tube settings).

The X-ray tube emits photons with a continuous background in the spectrum. Because the bias threshold depends on the photon energy, the count rate increases continuously with the bias current. The exact form of this increase depends on the details of the tube spectrum. As the high-energy part of the spectrum is reduced for low $V_A$, the count rate builds up slower with $I_b$ for lower $V_A$ and cannot just be scaled with a constant factor (shown with filled and empty symbols in Fig. 3(b)).

Measurements at higher temperatures (4-5 K) show that the count rate increases more steeply with $I_b$, and count rates up to $\approx 2 \cdot 10^4\,\text{s}^{-1}$ are observed at $T = 5\,\text{K}$. The increase of the count rate with temperature can be attributed to the temperature-dependence of the bias threshold, which is reduced at higher temperatures. Using the hot-spot model, a threshold of more than $220\,\mu\text{A}$ (for 50 keV) can be calculated [14] for $T = 4\,\text{K}$, which is right above the latching limit and might explain the elevated count rates at this temperature. At $T = 5\,\text{K}$, the corresponding thresholds are calculated to $110\,\mu\text{A}$ and $70\,\mu\text{A}$ for 6 kV and 50 kV photons, respectively, which is within the measurable bias region and which therefore explains the steep increase of the count rate with increasing $I_b$.

Note that detector TaN-B covers a significantly larger area than detector TaN-A, and a theoretical maximum count rate of $\approx 6 \cdot 10^4\,\text{s}^{-1}$ can be estimated for a non-latching X-SNSPD of the size of detector TaN-B with the same film properties.

### F. Dark-Count Rate Measurements

Despite latching at photon detection for high bias currents, we performed measurements on possible dark counts for $I_b / I_c \approx 92\%$ and $\approx 98\%$ at $T = 1.85\,\text{K}$ for the detectors TaN-A and TaN-B, respectively. No dark counts were recorded during more than 30 min measurement time, thereby limiting the dark-count rate to $< 5 \times 10^{-4}\,\text{s}^{-1}$. As explained in [1], negligible dark-count rates can be theoretically expected in principle for X-SNSPDs.

### G. Pulse Amplitudes

In Fig. 4(a) and (b) we plotted pulse-amplitude histograms for detector TaN-A (at $I_b = 54\,\mu\text{A}$) and TaN-B (at $I_b = 243\,\mu\text{A}$) for $T = 1.85\,\text{K}$, respectively, normalized with respect to the total number of counts. The amplitudes are much larger than for the Nb X-SNSPD [1] at $I_b = 0.61\,\text{mA}$ shown in Fig. 4(d) despite the smaller bias currents used, which is expected from the larger domain resistances. This magnitude of the amplitudes therefore allows for the recording of all detection pulses at $T = 1.85\,\text{K}$, and no pulses had to be cut off to eliminate noise signals.

However, for detector TaN-A, the expected large domain resistances must also lead to a reduction of the amplitude dependence on the photon energy spectrum as compared to the Nb X-SNSPD, because the pulse amplitude can only be significantly varied for domain resistances of the order of 50 Ω or smaller [1]. A weak amplitude dependence is visible only between $V_A = 6\,\text{kV}$ and 15 kV for this detector. This is in qualitative agreement with the reported 1-9 keV argon ion detection by a 10 nm thick and 800 nm wide NbN detector with comparable typical domain resistances [17], where no energy-dependence in the amplitude distribution was found.

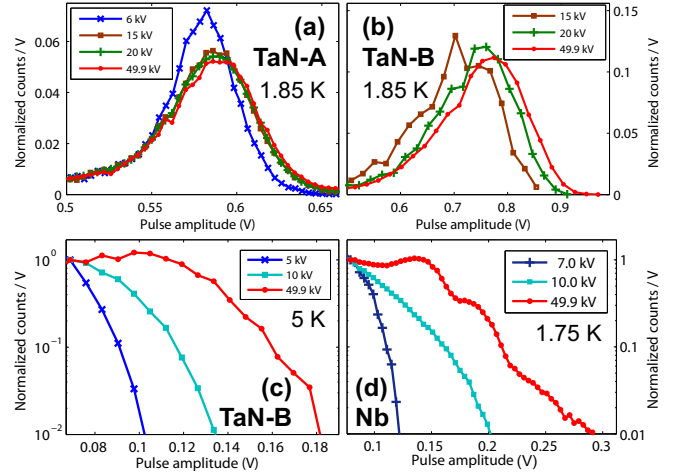

Fig. 4. Pulse-amplitude histograms for different $V_A$ for the detectors (a) TaN-A (at $T = 1.85\,\text{K}$ and $I_b = 54\,\mu\text{A}$) and (b) TaN-B (at $T = 1.85\,\text{K}$ and $I_b = 243\,\mu\text{A}$), normalized to the total number of counts. We also show a comparison for (c) TaN-B (at $T = 5\,\text{K}$ and $I_b = 121\,\mu\text{A}$) and (d) the Nb X-SNSPD from [1] (at $T = 1.75\,\text{K}$ and $I_b = 0.61\,\text{mA}$), here normalized relative to the counts at an amplitude of 69 mV and 79 mV, respectively.

Detector TaN-B shows larger amplitudes than TaN-A despite its smaller domain resistances, which is a consequence of the larger applied bias current. Furthermore, the amplitude distribution shows a stronger dependence on the photon energy spectrum than for detector TaN-A, as the domain resistances must be smaller. Below $V_A = 15\,\text{kV}$ and at $T = 1.85\,\text{K}$, the count rate is very small. However, as already shown in Fig. 3(b), it increases significantly at higher temperatures, and at $T = 5\,\text{K}$ measurable count rates are achieved even for $V_A = 5\,\text{kV}$ as shown in Fig. 4(c), as low-energy photons become detectable. However, the bias current needs to be reduced due to a lower critical current, and therefore certain pulses escape detection, as the discriminator level is set above the noise level. We faced the same problem for the Nb X-SNSPD [1] (see Fig. 4(d)), and we therefore normalized Fig. 4(c) and (d) with respect to the counts of a bin right above the noise level. At $T = 5\,\text{K}$, TaN-B shows a pronounced energy dependence of the amplitude distribution which is qualitatively very similar to that of the Nb X-SNSPD, where the amplitude distribution broadens along with a broader photon energy spectrum.

### H. Device Detection Efficiency

Using the Fe-55 X-ray source, we determined the DDE of detector TaN-A for 5.9 keV photons to 1.4%. Considering the absorptance [18], geometry and filling factor of the superconducting meander, a maximum DDE of 1.9% could in principle be achieved if all photons absorbed in the superconducting meander were counted. To achieve this value in the present detectors is not realistic, however, due to the nature of the energy-diffusion mechanisms involved [1].

## V. Conclusion

In conclusion, our results show that 100 nm thick TaN X-SNSPDs offer an improved detector performance as compared to 100 nm Nb X-SNSPDs concerning DDE, latching and pulse amplitudes. Ultrafast dark count-free soft X-ray single-photon detection is observed. Wide conduction paths allow for a certain energy-resolving capability in contrast to narrow conduction paths. We expect that X-SNSPDs from thicker TaN films should provide an even larger DDE. The latching tendency might be further reduced by a recently proposed biasing scheme [19].